\documentclass[preprint,journal]{vgtc}       

\ifpdf
  \pdfoutput=1\relax                   
  \usepackage{graphicx}                
  \DeclareGraphicsExtensions{.pdf,.png,.jpg,.jpeg} 
\else
  \usepackage{graphicx}                
  \DeclareGraphicsExtensions{.eps}     
\fi%

\graphicspath{{figures/}{pictures/}{images/}{./}} 

\usepackage{microtype}                 
\PassOptionsToPackage{warn}{textcomp}  
\usepackage{textcomp}                  
\usepackage{mathptmx}                  
\usepackage{times}                     
\usepackage{cite}                      
\usepackage{tabu}                      
\usepackage{booktabs}                  

\usepackage[T1]{fontenc}
\usepackage[utf8]{inputenc}
\usepackage{amsfonts}

\preprinttext{}

\onlineid{0}

\usepackage{amsmath}

\usepackage{xcolor}

\vgtccategory{Research}

\title{Mutual Scene Synthesis for Mixed Reality Telepresence}

\author{Mohammad Keshavarzi${\;  }^{1, 2}$ 
\and Michael Zollhoefer ${\;  }^{ 1}$ %
\and Allen Y. Yang ${\;  }^{ 2}$
\and Patrick Peluse ${\;  }^{ 1}$
\and Luisa Caldas ${ \; }^{ 2}$}
\affiliation{ ${}^{1}$ Reality Labs Research, Meta \\
${}^{2}$ University of California, Berkeley}

\authorfooter{
\item Corresponding author's email: mkeshavarzi@berkeley.edu 
}

\shortauthortitle{Firstauthor \MakeLowercase{\textit{et al.}}: Paper Title}

\abstract{Remote telepresence via next-generation mixed reality platforms can provide higher levels of immersion for computer-mediated communications, allowing participants to engage in a wide spectrum of activities, previously not possible in 2D screen-based communication methods. However, as mixed reality experiences are limited to the local physical surrounding of each user, finding a common virtual ground where users can freely move and interact with each other is challenging. In this paper, we propose a novel mutual scene synthesis method that takes the participants' spaces as input, and generates a virtual synthetic scene that corresponds to the functional features of all participants' local spaces. Our method combines a mutual function optimization module with a deep-learning conditional scene augmentation process to generate a scene mutually and physically accessible to all participants of a mixed reality telepresence scenario. The synthesized scene can hold mutual walkable, sittable and workable functions, all corresponding to physical objects in the users' real environments. We perform experiments using the MatterPort3D dataset and conduct comparative user studies to evaluate the effectiveness of our system. Our results show that our proposed approach can be a promising research direction for facilitating contextualized telepresence systems for next-generation spatial computing platforms.  
} 


\teaser{
  \centering
  \includegraphics[width=\textwidth]{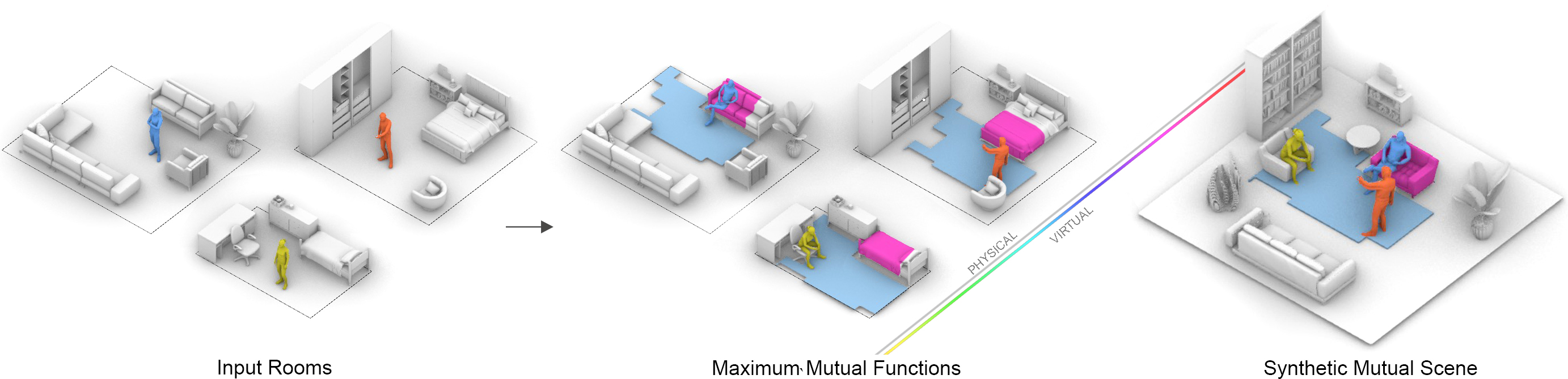}
  \caption{Mutual Scene Synthesis from three input rooms in a telepresence scenario. The system calculates optimal alignments to maximize mutual functional spaces, and furthermore generates a synthetic scene which incorporates the mutual functions with contextual placement of augmented objects.}
  \label{fig:teaser}
}

\renewcommand{\manuscriptnotetxt}{}

\begin{document}


\firstsection{Introduction}

\maketitle

Spatial Computing interfaces such as augmented reality (AR), virtual reality (VR) and mixed reality (MR) have become promising modalities for next generation computing platforms. While forming an expanding market in various applications in today’s technology space, they have shown potential success to promote remote social experiences \cite{beck2013immersive, gross2003blue,zendejas2013state,lorello2014simulation, buckley2014skillset}. However, spatial computing itself comes with spatial limitations. Such experiences are physically constrained by the geometry and semantics of the local user's environment where existing furniture and building elements are present \cite{narang2018simulating, Razzaque2001}. Contrary to 2D screens, where a rectangular screen region can always host digital content with possible overlay, 3D environments are occupied with physical obstacles and functional constraints. This results in complex, and many times non-convex, activity spaces available for virtual content augmentation. Therefore, surrounding physical boundaries would limit the free movement of users, prompting possible conflicts with the virtual experience. This limitation is elevated in multi-user remote scenarios, where identifying a common ground physically accessible for all participant becomes challenging. Carrying out such procedure is difficult by the users themselves, especially if they are unaware of the spatial layout of other participants' physical environment. \cite{keshavarzi2020optimization}.

\begin{figure*}
  \includegraphics[width=2\columnwidth]{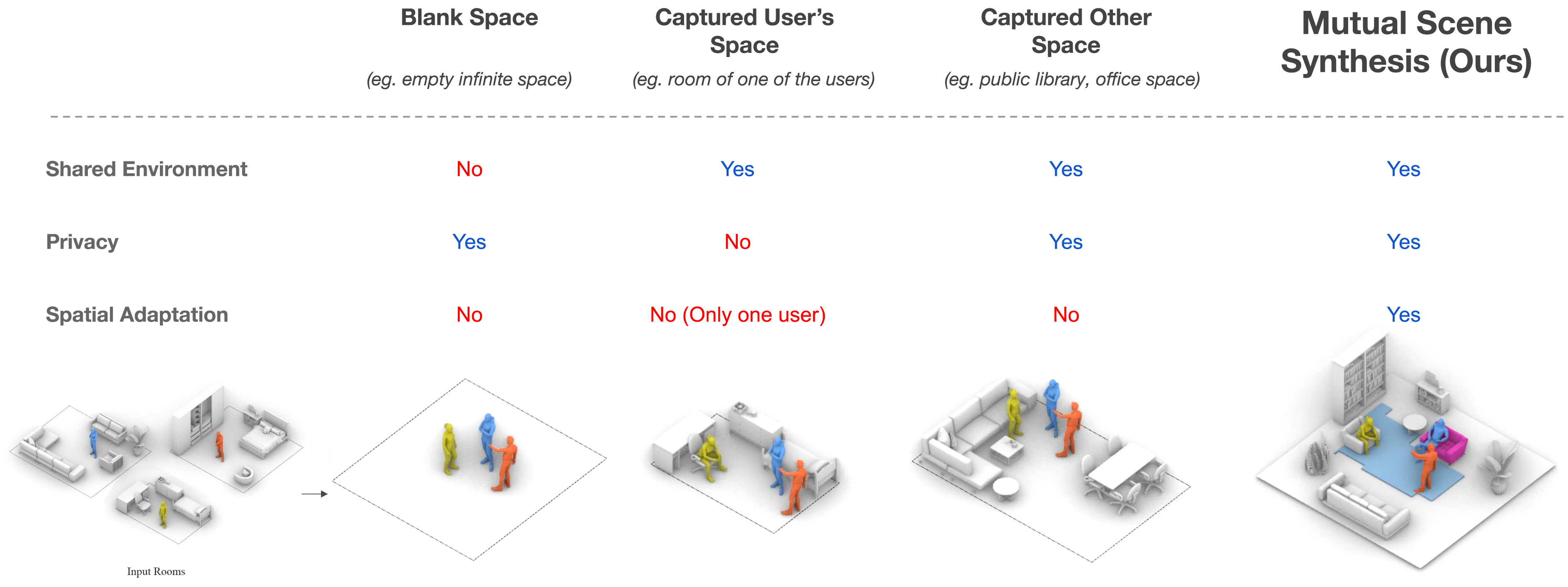}
  \caption[Graph Options]{Comparison between various telepresence scenarios. Our proposed mutual scene synthesis system can allow remote users to share a mutual environment while maintaining privacy and spatial adaption of their local physical environment.}~\label{fig:tel_scen}
\end{figure*}

Hence, a major challenge in developing telepresence systems is how to align and map virtual avatars within a target space, while addressing the spatial constraints of each user within their own local environments \cite{keshavarzi2020optimization}. Prototypes of high-fidelity telepresence systems \cite{lombardi2018deep,ma2021pixel,richard2021audio} could avoid this challenge by placing remote users in an empty virtual space exclusively defined for the task. Considering natural locomotion as a key aspect of maintaining high-fidelity experiences, users can therefore only perform interaction and navigation tasks within their own boundaries. However, the method lacks rendering a mutual environment and does not hold spatial correspondence with its local surroundings for each participating user, limiting free body movement the use of mixed reality 
features such as pass-through objects and preventing the ability to interact with existing physical entities within the telepresence experience.

Since shared mutual environments can potentially play an important role in increasing productivity and social engagement. There has been a large body of literature focusing on capturing surrounding environments that can be utilized in spatial computing applications \cite{mildenhall2020nerf,tewari2021advances,su2021nerf,martin2021nerf}. 
Such captures can be used as a spatial background for telepresence avatars while matching their environmental lighting for additional photo-realism.
However, attempting to capture detailed information from personal spaces can potentially cause privacy concerns and may unwillingly expose socioeconomic information of individuals during a telepresence call.
Capturing public spaces (such as office spaces, conference rooms, or cafes) can also be integrated within the telepresence environment. Yet, this approach also lacks spatial customization and interaction with the physical environment itself, isolating the experience to predefined spaces or calibrated functions.

In this paper, we focus on the spatial correspondence between the virtual and physical space as an effective feature for enhancing virtual experiences. Studies have shown that users of immersive experiences report a higher sense of presence when a match between proprioception and sensory data is achieved \cite{slater1993representations}. It is due to this match that natural locomotion has been shown to be superior to other navigation methods such as teleportation, flying or utilizing game controllers \cite{usoh1999walking}. In addition, from a safety perspective, as users of immersive environments are visually detached from their surrounding physical spaces, various techniques are utilized to inform the user of their physical surrounding, or deter them to prevent physical collisions. Alternative approaches aim to to generate a virtual experience that map to physical elements of the user's surrounding. For instance, a wall in the physical space would render as a barrier in the virtual environment, or a chair in the virtual environment would be sittable in the physical world. With all its potentials, such techniques do not extend to multi-user scenarios and cannot generate virtual environment that are adaptable to all participant's physical spaces.

Motivated by these challenges, we propose a contextual Mutual Scene Synthesis (MSS) system for spatial computing telepresence scenarios. Given a set of captured rooms, our proposed system generates a synthetic virtual scene that holds maximum functionality between the captured rooms and corresponds to their individual layouts. Users can safely navigate within the synthetic scene with natural locomotion and interact with mutual furniture that will have a physical correspondence
in their surrounding local environment. Our work builds on an emerging body of literature on scene synthesis, while taking advantage of the works done in mutual spatial alignment for telepresence scenarios. As illustrated in Figure \ref{fig:tel_scen},  when compared to alternative environment generation for telepresence scenarios, our mutual scene synthesis system enables a shared environment while maintaining privacy and spatial correspondence for each of the participants. We believe that utilizing this method can potentially facilitate spatial adaptations of next-generation computer-mediated communication platforms in spatial computing.

\section{Related Work}
\subsection{Collaborative Telepresence Systems}
Telepresence allows changing the state of one's sense of presence from a physical location to a target remote environment without requiring the physical body to relocate to the target environment \cite{fadzli2020review, fairchild2016mixed}. A large body of work has explored how collaborative human-based telepresence can be achieved by capturing a region of each participants body and space and projecting it to the target environment. Systems developed by \cite{wen2000toward, gross2003blue, kuster2012towards, beck2013immersive, zhang2013viewport,benko2012miragetable} are examples of such efforts where participants and a limited range of their surrounding spaces are continuously captured using a cluster of registered depth and color cameras. Recent work of \cite{lawrence2021project} in Project Starline takes this approach one step closer towards a high fidelity co-presence experience. The bi-directional system is able to capture audiovisual cues such as stereopsis, motion parallax, and spatialized audio; while enabling high resolution communication cues such as eye contact and body language.

However, window-based telepresence systems limit the participant's ability to access each others spaces. Users are spatially disconnected from each other and interaction occurs through a audiovisual window acting as a barrier. The importance of free-form user movement and the ability to preserve mobility-based communication features in the context of co-presence has been studied in the work of \cite{bardram2005activity,keshavarzi2019affordance,luff1998mobility, caldas2019design}.  Alternatively, research in room-based telepresence systems has gained major momentum in recent years, allowing bilateral tele-presence between participants, where participants share a common virtual ground. Work of \cite{Orts-Escolano2017} allows a remote user to be captured and rendered into a local user's space via an augmented reality head-mounted device (HMD), providing the feeling that the remote user is present in the local user's space as well. Such an approach is also seen in \cite{maimone2013general, fairchild2016mixed}, where the remote and local users do not share the same room layout, but are calibrated in order to provide the required mutual virtual ground between users. \cite{stotko2019slamcast} enables mutual ground sharing by capturing the local space of one of the participants, and streaming the data to a limited number of remote users. Recent work of Codec Avatars \cite{lombardi2018deep,ma2021pixel} implemented as a decoder network of a Variational AutoEncoder (VAE) demonstrates how high-fidelity animatable human head models can be captured and later rendered in real-time via spatial computing HMDs. The 6 degree of freedom HMDs allow all participants to freely walk within their local environment, while experiencing co-present with other participants' avatars within the virtual or mixed reality environment. 

\begin{figure*}
  \includegraphics[width=1.95\columnwidth]{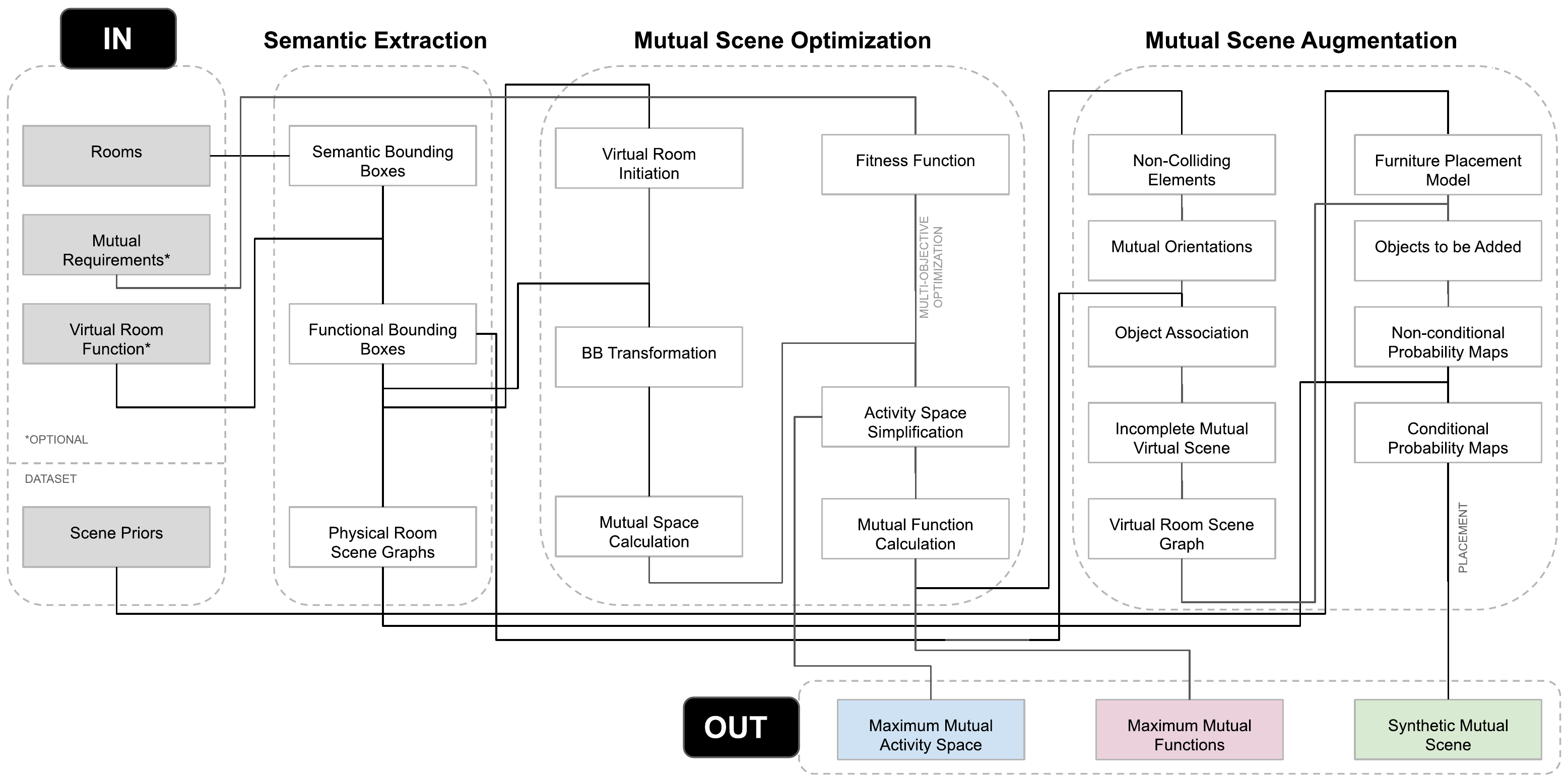}
  \caption[Graph Options]{General workflow of the Mutual Scene Synthesis (MSS) system.}~\label{fig:GenWorkflow}
\end{figure*}

\subsection{Mutual Spatial Alignment}
Mapping virtual avatars within a shared target space, while addressing the spatial constraints of each user within their own physical environment is considered an open challenge for next-generation mixed reality telepresence platforms \cite{keshavarzi2020optimization}.
Previous work has focused on methods to create mutual grounds and understand user preferences for different types of mutual ground generation. \cite{sra2018your} designed three mapping models (scale, kernel, and overlap) for aligning simple rectangular play area spaces. They further conducted extensive user experiments to evaluate participants' sense of co-presence. The work of \cite{Lehment2014,congdon2018merging} discussed methods that resembled our module the most. The systems there aimed to optimally map remote environments to maximize user activity space and minimize obstacle discrepancy. In contrast, our mutual space module offers a multi-function optimization workflow, allowing a user in the loop to define weights and constraints for multiple mutual function rooms. For example, instead of finding the maximum walkable space, the user can choose to have a smaller walkable space while maintaining a mutual workable space with other participants. Our method also utilizes an evolutionary optimization algorithm, more suitable for processing multiple input spaces instead of just two spaces.
Another previous study \cite{keshavarzi2020optimization} calculated maximum mutual space alignment, and recommended furniture reconfiguration strategies that minimize the effort to move objects while increasing the mutual space.

Another feature of our work is to generate a virtual experience that maps to a physical object in the user's environment. Such an approach has been widely explored in work of \cite{cheng2017sparse, cheng2017mutual, cheng2018iturk}. In \cite{cheng2015turkdeck}, a group of real people are instructed to dynamically change a physical environment of props to provide haptic feedback for a user in VR. Our work however aims to generate a virtual experience that correspond to the natural livable personal environment of the user, instead of calibrating props within the physical environment. This is seen in the work of \cite{lindlbauer2018remixed} where a live 3D reconstruction from external depth cameras is utilized to allow modification of the scene, including adding custom virtual objects. In \cite{sra2016procedurally}, after identifying obstacles and walkable areas of physical space, the authors use a procedural model to generate a planar walkable space within a predefined virtual environment. 
Both systems mentioned above are limited to a single user space, while our proposed method is designed for telepresence applications, thus aims to provide spatial correspondence to multiple spaces. Moreover, while our method also utilizes a procedure model as part of the synthetic scene initialization, we further take advantage of learning-based deep networks, trained on real-world scenes to fully synthesize the virtual scene.

\subsection{Scene Synthesis}

Scene synthesis aims to generate a plausible scene layout while satisfying both functional and aesthetic criteria  \cite{zhang2019survey}. Early work focused on hard-coding rules, guideline and grammars, commonly referred to as procedural modeling \cite{bukowski1995object, xu2002constraint, germer2009procedural}. Rules could be extracted through layout manuals and interviewing professionals as seen in the work of \cite{merrell2011interactive, Yu2011}. Such an approach was extended in \cite{yeh2012synthesizing} by attempting to synthesize open world layouts with hard-coded factor graphs. Example-based synthesis was later introduced by \cite{Fisher2012} where they developed probabilistic models on Bayesian networks and Gaussian mixtures using an example set of scenes. Work of \cite{kermani2016learning} synthesized a full scene iteratively by adding a single object at a time. Similar to our approach, their model trained through pairwise and higher-order object relations, but was only limited to object relations and did not capture room-object relationships.
Work of \cite{Liang2018, Liang2017,fu2017adaptive} take room functions also into account during synthesis, something we optionally allow users of the MSS system to define.


In \cite{wang2018deep,ritchie2019fast}, sequential scene synthesis of partially completed scenes takes place by learning from top-down images of 3D scenes as priors. This approach is improved in \cite{wang2019planit}, where a combination of object-level and high-level separate convolutional networks are utilized to address constrained scene synthesis problems. 
Work of \cite{paschalidou2021atiss, wang2021sceneformer} take advantage of auto-regressive transformers within their scene synthesis architecture. Contrary to previous work, where separate models are trained to generate object attributes such as category, location, etc., while \cite{paschalidou2021atiss} requires a single-model training procedure to predict
all attributes. Our scene synthesis module comes close to the work of \cite{zhou2019scenegraphnet,keshavarzi2021contextual}, which utilizes a scene graph representation to describe a wide variety of object-object and object-to-room relationships, and tend to conduct constrained scene synthesis by learning from graph priors. Our main difference between the prior scene synthesis work mentioned above is our scene synthesis module is conditional, aiming to place objects in a plausible manner in the virtual scene while attempting to correspond to one or more physical objects present in target remote rooms.

\begin{figure}
  \includegraphics[width=1\columnwidth]{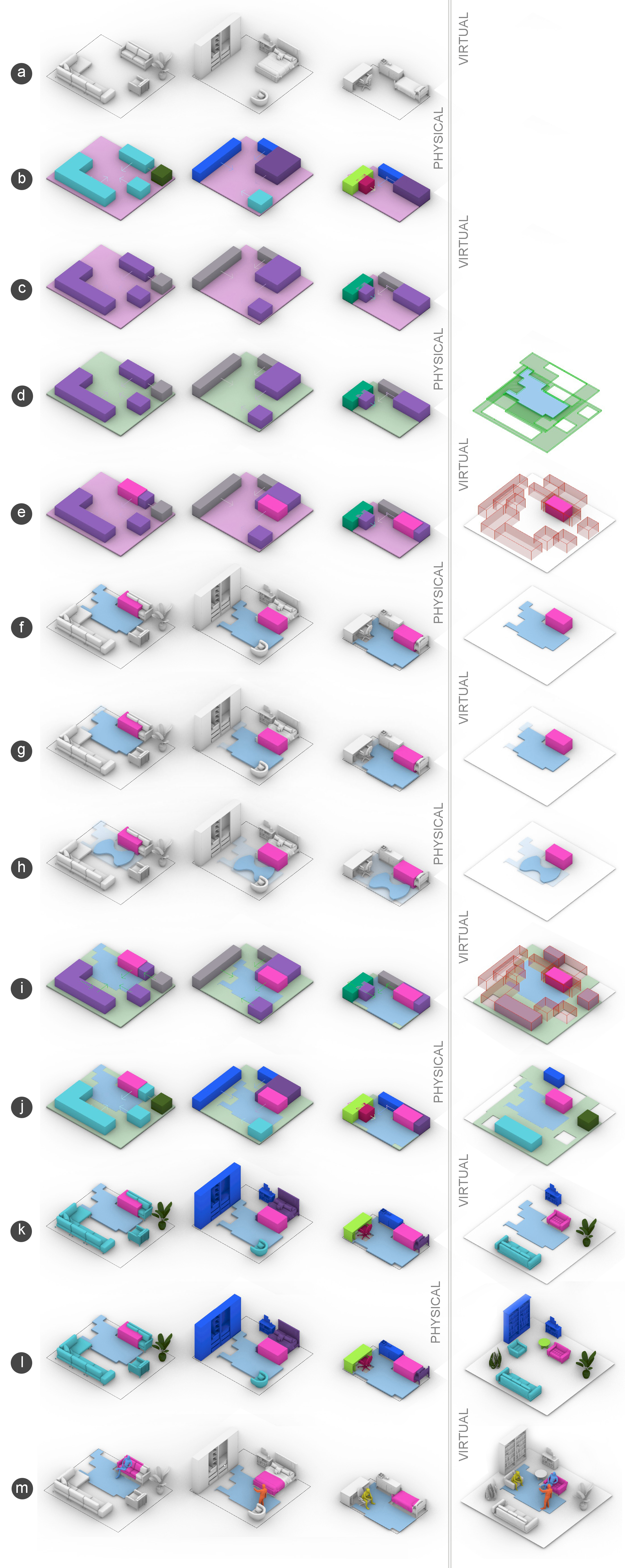}
\caption{A step by step example of each of the modules in our Mutual Scene Synthesis (MSS) framework. General components include: (b) input rooms (b-c) semantic extraction (d-h) mutual scene optimization (i-l) mutual scene augmentation (m) synthetic mutual scene output. }~\label{fig:full_workflow}
\end{figure}

\section{System Overview}

Figure \ref{fig:GenWorkflow} shows the workflow of our proposed system. The system takes the collection of rooms of the remote participants as input and generates a synthetic virtual scene with maximum mutual functions corresponding to the input rooms. Our proposed system consists of three main components: (i) Semantic Extraction: where a simplified semantic scene graph representation of the room is extracted; (ii) Mutual Scene Optimization: where the maximum mutual functions are calculated between the input rooms; and (iii) Mutual Scene Augmentation: where conditional scene augmentation is conducted using deep neural network models trained via scene priors. 
Figure \ref{fig:full_workflow} illustrates various steps of the mutual scene synthesis system using an example set of rooms. 
%
%
We discuss the details of each component in the following sections.

\section{Scene Representation\label{sec:sceneRep}}
\subsection{Rooms and Objects}
In this paper, we define the room space $R$ as an orthographic projection of its 3D geometry on the $(x,y)$-plane. We denote the $k$-th object (e.g., a chair or a bed) in $R$ as $O_{k}$. The collection of $n$ objects in $R$ is denoted as $\mathcal{O}= \{O_1, O_2,... O_n\}$.  $B(O_k)$ represents the bounding box of the object $O_k$. Every object $O_k$ has a label to classify its type. As our work requires multiple user spaces, we define for each user $i$ their own room space expressed as $R_i$ and the $k$-th object in $R_i$ is denoted as $O_{i,k}$. Hence, the collection of all $n_i$ objects in $R_i$ is denoted as $\mathcal{O}_i= \{O_{i,1}, O_{i,2},... O_{i,n_i}\}$. Finally, we define the area function as $K(O)$.

We also differentiate between physical rooms and the virtual room in our notation. A virtual room is considered a room fully or partially rendered in mixed reality. We denote the virtual room as $R^{\prime}$ and the virtual objects as $O^{\prime}$. In addition, we introduce a distance function $\delta(a, b)$ as the shortest distance between $a$ and $b$ objects. For example, $\delta(B(O_{k}),\dot{R})$ is the shortest distance between the bounding box of $O_k$ and the center of the room $R$.

\subsection{Semantic Scene Extraction}
We consider the input to our system to also include semantically labeled bounding boxes. Semantic bounding boxes can either be defined manually by the user in MR \cite{saran2019augmented}, or be an output of automated semantic segmentation systems such as \cite{Qi2016,Liu2018,Armeni2016}. Therefore, every object $O_k$ has a label to classify its functional type (see Figure \ref{fig:full_workflow}. b). Furthermore, we define functional categories to describe objects and spaces with similar functional types. In our current implementation, each $R_i$ can hold various functional categories of walkable ($W_{i}$), sittable ($S_{i}$) and workable ($T_{i}$) spaces. Walkable spaces consist of the area of the room in which no object located within a human user's height range is present. In walkable spaces, user movement can be performed freely without any risk of colliding with an object in the room. We calculate the available ($W_{i}$) for room $R_i$ simply as follows:

\begin{equation}
W_{i} =  R_i - \bigcup_{k=1}^{n_i}O_{i,k}.
\end{equation}

Sittable and workable spaces correspond directly to a group of objects within a room. For example, chairs, sofas, beds, stools, etc.~are all considered to have a sittable functionality. Objects such as desks, tables, etc. are considered workable functions. For $R_i$ we have $S_{i} = \{ {O^s}_{i,1}, {O^s}_{i,2},... {O^s}_{i,n_i} \}$, where ${O^s}_{i,k}$ is considered an object in $R_i$ which holds a sittable function. A similar notation is true for workable function groups defined as $T_{i}$. Figure \ref{fig:full_workflow}. c) illustrates how functional semantic segmentation takes place in the input rooms.

\subsection{Semantic Scene Graphs}\label{sec:graphExtraction}

\begin{figure}
  \includegraphics[width=1\columnwidth]{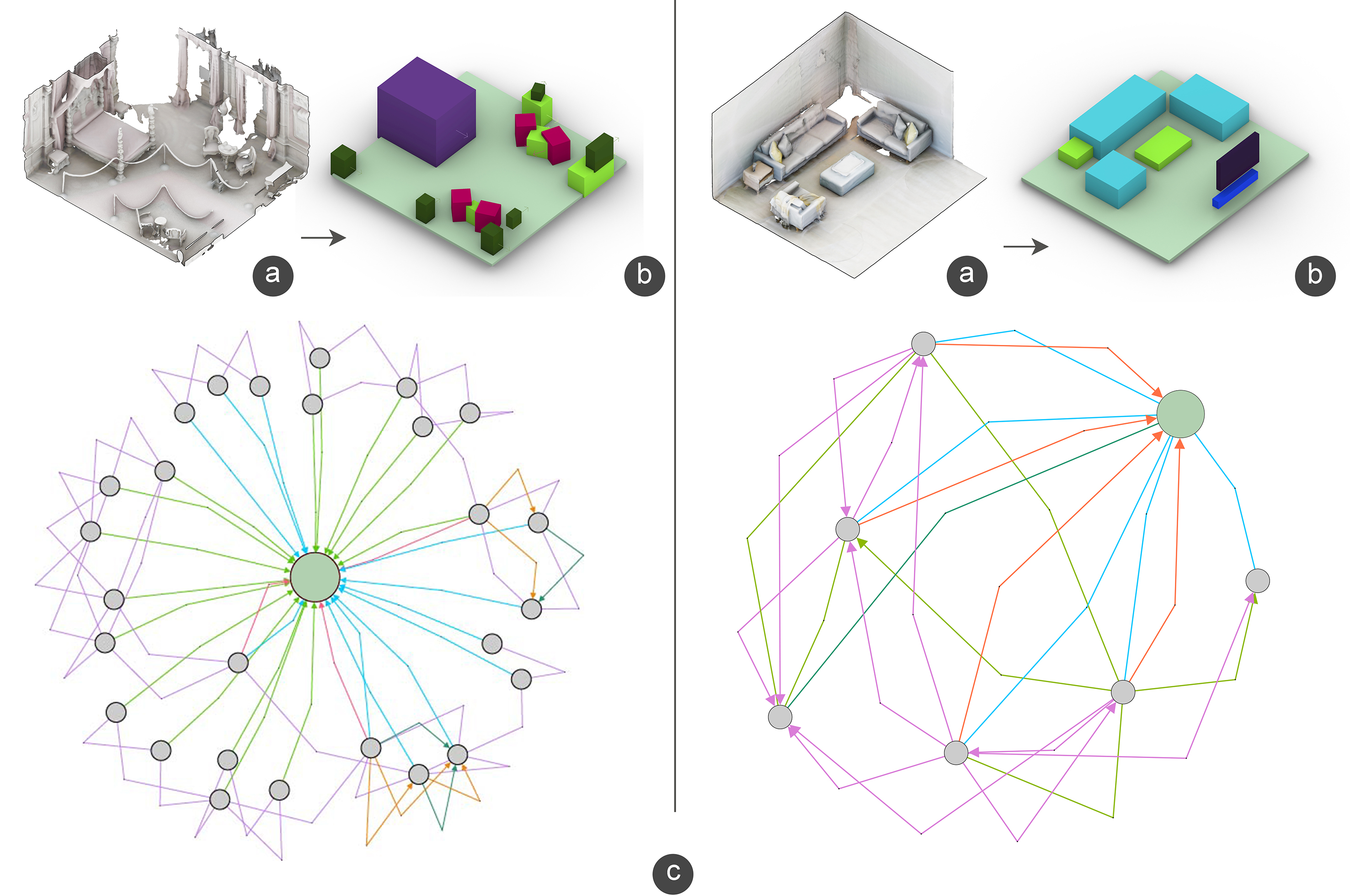}
  \caption{Examples of semantic scene graph extraction. (a) input room; (b) semantic segmentation (c) collection of semantic scene graphs. Semantic scene graphs represent pair-wise relationships between objects and the room.}~\label{fig:semnaticgraph}
\end{figure}

To capture contextual topologies between objects of a scene, we represent rooms via semantic scene graphs. We utilize homogeneous scene graphs via the spatial relationship introduced in \cite{keshavarzi2020scenegen} to construct those scene graphs. Nodes in the scene graph represent objects, object groups, and the room; and edges represent the spatial relationships between the nodes allowing to describe the pair-wise topologies of objects and their relationship with the room. In the proposed scene graph representation, an explicit extraction of (a) positional and (b) orientational relationships take place by modeling descriptive topologies that are commonly utilized by architects and interior designers to generate spatial functionalities in a given space. Figure \ref{fig:semnaticgraph} illustrates an example of semantic scene graph collections for two input scenes. Note that each edge color corresponding to a spatial relationship represents a separate scene graph. Such a representation allows contextual scene augmentation to be utilized for an incomplete scene by training with previous scene graph priors. Further details of the scene augmentation process is discussed in Section \ref{sec:sceneAug}.

\section{Mutual Space Optimization}
The goal of this module is to calculate optimal functional mutual spaces between participants by aligning the participants local spaces within the virtual environment. The mutual functional spaces generated in the virtual environment correspond to real-world functions in remote participants within their local environments. Such spaces are calculated by finding the optimal transformation function for each space to maximize the intersection of all spaces. We consider an immersive experience where there are $m$ subjects and therefore $m$ room spaces $(R_1, R_2, \cdots, R_m)$,
respectively. Then, in the $(x,y)$-coordinates, we define a rigid-body motion 
in $\mathbb{R}^2$
as $G(F, \theta)$,
where $\theta$ describes a translation and a rotation. 

To maximize the mutual walkable space, we apply one $G(W_i, \theta_i)$to each individual walkable space $W_i$ for the $i$-th user. The optimal rigid body motion then maximizes the area of the interaction space:
\begin{equation}
    (\theta_1^*, \cdots, \theta_m^*)=\arg\max K(\bigcap_{i=1}^{m} G(W_i, \theta_i)).
\label{eq:optimal-rigid-body-motion}
\end{equation}

Hence the maximal mutual walkable space can be calculated as
\begin{equation}
    M_W(R_1,\cdots, R_m) = \bigcap_{i=1}^{m} G(W_i, \theta_i^*)
    \label{eq:max-standable-space}
\end{equation}

\subsection{Mutual Functions}
Similar to walkable spaces, our system calculates mutual areas of remaining functional categories namely mutual sittable ($M_S$) and mutual workable ($M_T$) spaces. The main difference between mutual walkable spaces and mutual function areas is that mutual functions require pose estimation. We use the following heuristic to define the pose of the calculated mutual functions: If the objects constructing the mutual function share the same pose direction, the mutual function area would also hold that pose direction. Else, the mutual function would be facing the center of the room. In Figure \ref{fig:full_workflow}. e) mutual function optimization takes place, classifying a section of the sittable area of the sofa in $R_1$, and the bedspace in $R_2$ and $R_3$. Due to the fact that the pose of the bed in $R_3$ differs from the other corresponding sittable functions in $R_1$ and $R_2$, the resulting sittable space pose is calculated towards the center of the room. 

\subsection{Geometry Simplification} In certain scenarios, participants of a telepresence experience may require the mutual activity space to comply to a minimum area or hold a certain shape. Games for instance may require users to hold a safe play area, often being a quadrilateral or circular space to avoid physical conflicts. Another possible example is when users are surrounding and inspecting an object, and the activity space is preferred to be a circular shape with the object placed in the center. 
To this extent, mutual spaces solely calculated based on maximizing functional areas may hold non-convex peninsula-like geometry, which can become inaccessible for various activities. For instance, the mutual walkable space calculated in Figure \ref{fig:full_workflow}. f) holds areas which a regular human body cannot perform free body movement without colliding with the boundaries of the space.

To address such scenarios, we add two optional post-processing modules to our workflow to generate safe activity spaces which allow: (a) simplification of the resulting mutual geometry to exclude peninsula-like areas (Figure \ref{fig:full_workflow}. g); and (b) calculation of the largest custom activity shape inscribed in the mutual geometry (Figure \ref{fig:full_workflow}. h). For simplification, a double-stage offsetting procedure takes place. In the first stage an inward offset with a distance of $\epsilon$ is conducted from the bounding polygon of the mutual space. Edges with more than two intersections are removed, resulting in a simplified inward offset of the shape. An outward offset with distance $\epsilon$ is followed as the second stage, generating a shape similar to the initial shape with excluded peninsula-like areas. The size of $\epsilon$ can be defined based on the activity. For instance, intense gaming applications that involve a high level of free-body movement would require a larger $\epsilon$ than a normal natural locomotion activity.

For calculating the largest inscribed custom activity space $L_S$, we define a rigid body function in $\mathbb{R}^2$ as $J(L, \theta, s_x, y)$, where $L$ is the custom activity shape, $\theta$ describes a translation and a rotation, and $s_x,s_y$ are scale factors applied to $L$ is the $x,y$ direction respectively. We run the following optimization:

\begin{equation}
\begin{aligned}
(\theta^*, s_x^*, s_y^*)= \arg\max {} & (K(J(L, \theta, s_x, s_y)\bigcap  {M_W})\\
& - K(J(L, \theta, s_x, s_y)\bigcap  {M_W}^\prime))
\end{aligned}
\end{equation}

Where ${M_W}^\prime$ is the inverse of the mutual space ($R_i  - M_W$).
Hence, the largest custom activity space is calculated as:
\begin{equation}
L_S = J(L, \theta^*, s_x^*, s_y^*)
\end{equation}
Figure \ref{fig:full_workflow}. h) shows an example of a optimization achieved to find the custom polygon inscribed in the mutual boundaries

\begin{figure*}
\includegraphics[width=2\columnwidth]{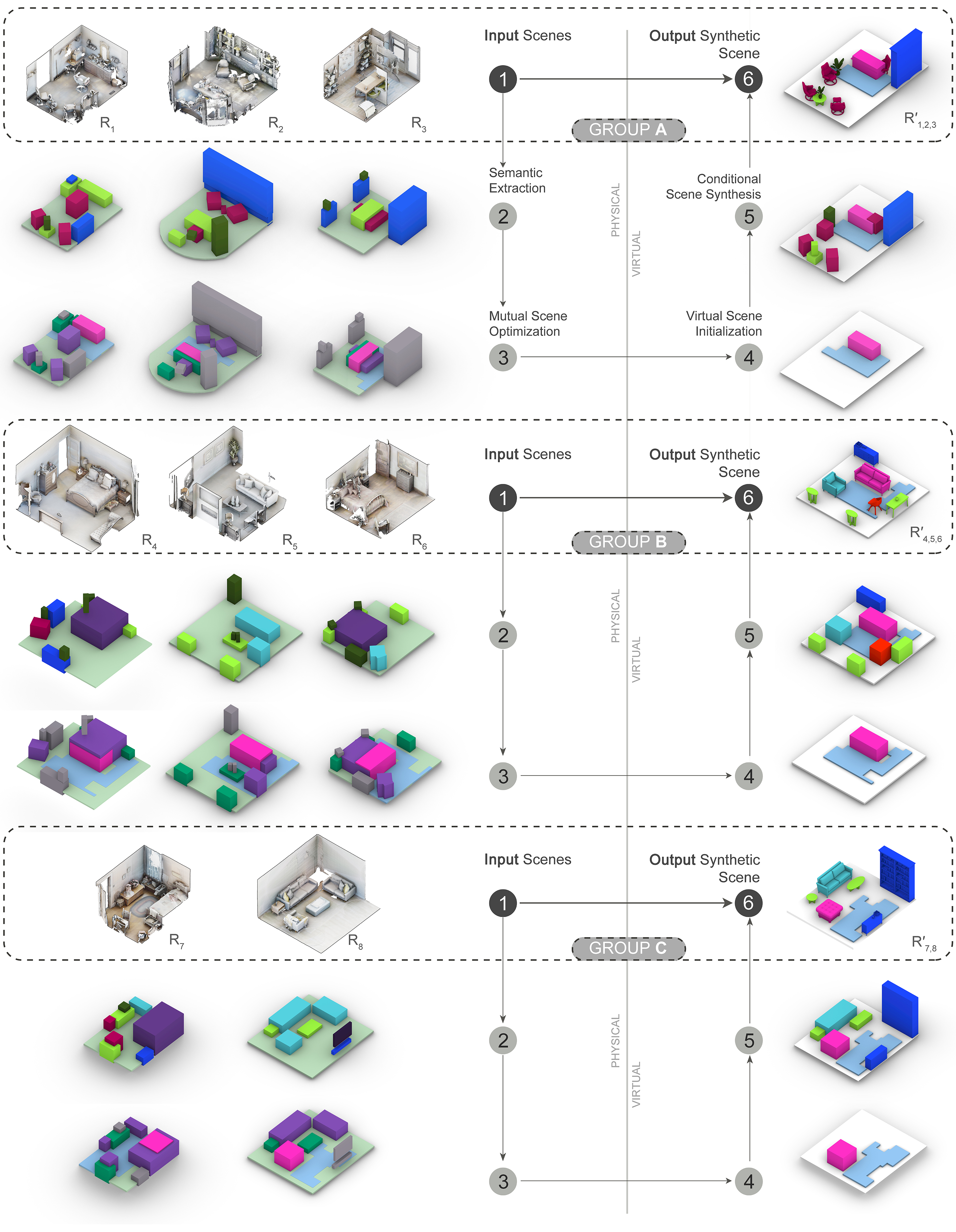}
 \caption{Results of the MSS system on MatterPort3D dataset examples.}~\label{fig:3DExamples}
\end{figure*}

\subsection{Optimization}
Considering various user-in-the-loop scenarios, optimizations can be defined as single objective or multi-objective problems. In telepresence settings that require only one mutual function type to be maximized, a single objective optimization is utilized to find the required transformation parameters for the room alignment. Alternatively, multiple functions can also have various weights and constraints (such as minimum sittable or workable area), reducing the dimension of the optimization to a single-objective function. This approach was utilized in the mutual function optimization in Figure \ref{fig:full_workflow}. d) and \ref{fig:full_workflow}. e) where a minimum projected mutual sittable space of 1$m^2$ was defined as a constraint, while maximization of the walkable space took place. Users can also be involved within the workflow for multi-function scenarios, where a set of solutions representing the Pareto frontier of the multi-objective optimization would be presented to the user. After considering trade-offs, the user can choose which spatial configuration would be more suitable for their activity.

\section{Mutual Scene Augmentation}

After calculating the optimal alignment of target rooms, we aim to synthesize a new virtual scene which would incorporate the mutual spaces and provide a plausible virtual environment spatially corresponding to all target users. The Mutual Scene Augmentation process consists of two modules: the first module utilizes a procedural grammar for initializing the scene, followed by the second module that uses scene priors for conditional scene synthesis.

\subsection{Procedural Initialization}
As a first step of the virtual scene initialization, we define the base floor of the synthetic room as the smallest circumscribed rectangle of the union of all the aligned rooms. This would guarantee users to access all their available physical space within the virtual environment. Furthermore, we populate the synthetic room with non-colliding elements of each local space (Figure \ref{fig:full_workflow}. i) An object is considered non-colliding if (a) its transformed projection on the $(x,y)$ plane does not collide with another room's walkable space and (b) its transformed position does not collide with another transformed object in the mutual alignment. Adding non-colliding objects to the virtual scene would add an additional visual barrier to prevent a physical collision for the user holding the object in its local space. Once the bounding boxes of mutual functions and non-colliding objects are calculated, the system takes on the task of associating each calculated bounding box to a function type and furthermore to a designated mesh. During the object association step (Figure \ref{fig:full_workflow}. j), the function of the mutual room determines what objects should be placed in the scene synthesis step. The synthetic room function is an optional input given by the user of the system. If no input is given, the system uses the most repeated room function in the target set. If no majority is present, one of the room functions would be assigned randomly.


\subsection{Conditional Scene Synthesis via Priors}\label{sec:sceneAug}
As a final step, we use a deep-learning model to complete the room with additional furniture. The furniture is augmented in a conditional manner, taking into account relative furniture arrangements of input rooms. We utilize a modified version of GSACNet \cite{keshavarzi2021contextual} for the conditional scene augmentation process. GSACNet combines graph attention, siamese, and autoenoder networks to perform iterative scene synthesis for new or constrained scenes. For the training process, in order to achieve robust results with limited scene priors, we propose utilizing the parametric data augmentation method introduced in \cite{keshavarzi2020genscan}. In this method, after building parametric floorplans of the rooms, boundary geometry and their adjacent furniture are constantly permutated while maintaining a set of functional constraints within the room.  

For the scene augmentation process, the system initially samples points uniformly in the $(x, y)$-plane. Each point is considered as the center of possible placement for a target object $\dot{O_k}(x,y)$ on the ground floor plane, and forms its corresponding scene graphs discussed in Section \ref{sec:graphExtraction}. Next, the system passes feature vectors associated with nodes in the scene through an initialization neural network followed by a respective graph attention layer. Messages passed to the node associated with the object's furniture type are extracted and concatenate the messages per each scene graph with the summary vector. Furthermore, the concatenated vector is projected via a 4-layer network into a space such that data points representing plausible placements are clustered together while data points representing implausible placements are separated from the cluster. Finally, we output a probability of plausible placement $P$ using the reconstruction error produced by the an autoencoder. Studies have shown autoencoders to perform well as anomaly detectors \cite{zong2018deep}. In our scene synthesis system, there is a model per furniture group. The system trains each model using two separate training phases. In the first phase, the initialization, scene graph extraction, and project modules are trained as one large siamese network, with a siamese network projection module. In the second phase, the outputs of the first training process are used as input and train the autoencoder module alone.

In a conventional scene synthesis scenario, $\dot{O_k}(x,y)$ is placed in the location with the highest $P$. Instead, in our approach, we add an additional conditional module to allow contextual placements to take into account the arrangement of the real-world user target scenes in addition to the generated synthesized scene. The conditional module takes the $n$ top samples of $P$ and sorts them based on their distance to the closest object in the same functional type from all the input physical spaces. In simple terms, from the placements that the scene synthesis module considers plausible, the system chooses the final placement based on its proximity of real world objects in one (or more) of the real-world user spaces. Such an approach would assist the scene augmentation process to place objects closer to where they are in the real-world, potentially corresponding to one of the target room furniture arrangements. Hence, slightly contrary to conventional scene synthesis systems, our proposed approach populates the virtual scene by placing objects close to their real-world setting while being contextually relevant to the mutual virtual scene(Figure \ref{fig:full_workflow}. l) . 

\section{Experiments}
\subsection{Synthetic Generation via Real-world Datasets}\label{sec:sceneGen_exp}
To evaluate how our proposed mutual scene synthesis system performs with various room types and different spatial organizations, we utilize available 3D datasets from captured real-world scenes as case studies. We use the Matterport 3D {\cite{Chang2018}} dataset and sample subsets of varying size and functions of rooms, to observe how mutual spaces are optimized and the corresponding synthetic scene is generated. Matterport 3D is a large-scale RGB-D dataset containing 90 building-scale captured models. The dataset consists of various building types with diverse architecture styles, each including numerous spatial functionalities and furniture layouts. Human-defined annotations of building elements and furniture are provided with surface reconstructions as well as 2D and 3D semantic segmentation. We utilize this data for the semantic segmentation process. In addition, we exclude spaces that are not typically used for telepresence spaces (bathroom, small corridors, stairs, closet, etc.). 

\begin{figure*}[t]
  \includegraphics[width=2\columnwidth]{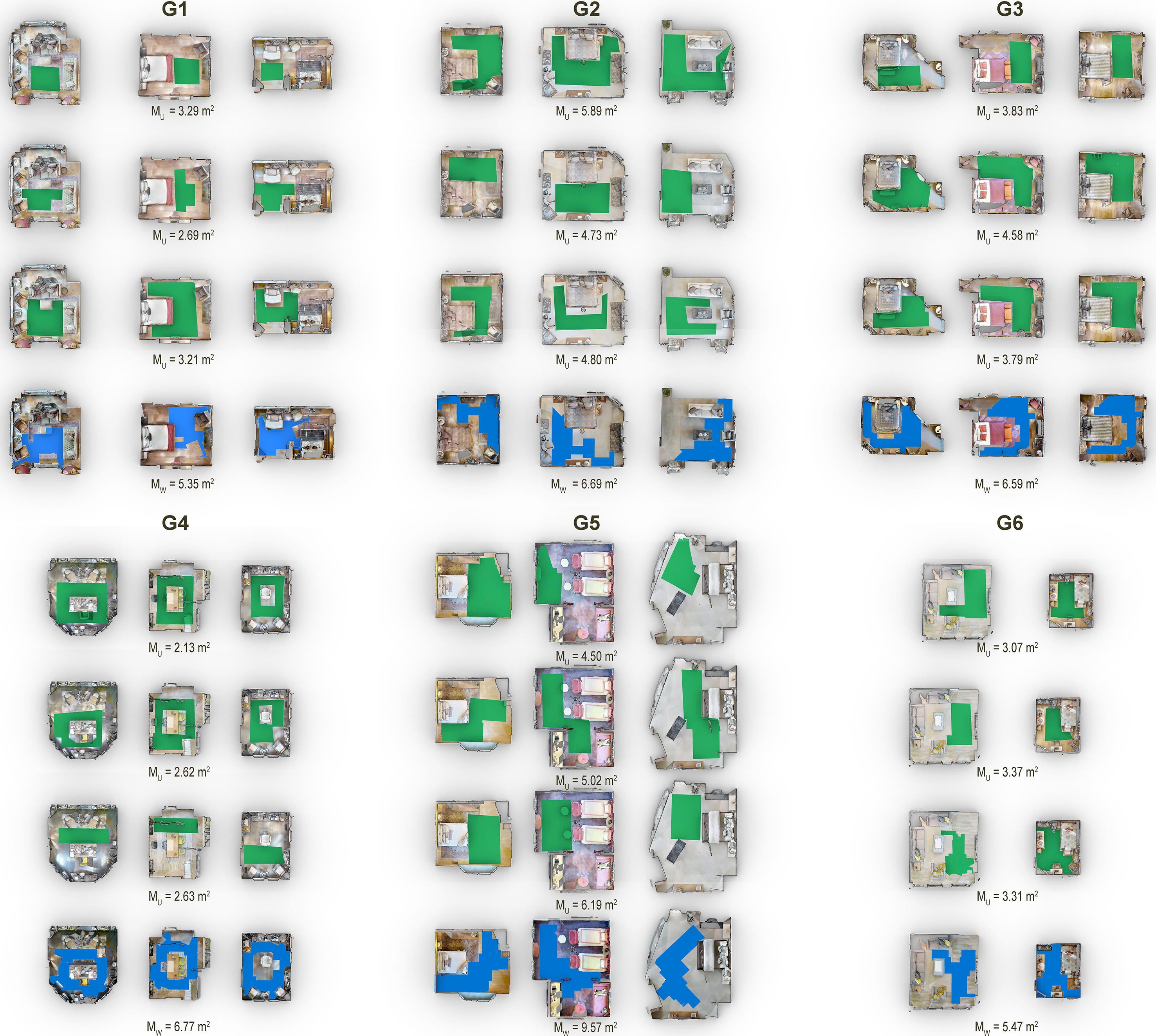}
  \caption{Results of the comparative user study showing top 3 human classification of walkable spaces ($M_U$) compared with the MSS system ($M_W$).}~\label{fig:userStudies}
\end{figure*}

For the mutual function optimization procedure, we utilize a Strength Pareto Evolutionary Algorithm 2 (SPEA 2) \cite{zitzler2001spea2} algorithm to calculate the maximum mutual functions between the rooms. We use a population size of 100, mutation probability of $10\%$, mutation rate of $50\%$ and crossover rate of $80\%$ for our search. As our solution integrates a evolutionary search, we expect the result to gradually converge to the global optimum. We stop the optimization after $80$ generation runs. Room translations are executed in 10cm steps in the $(x,y)$ plane and $15^\circ$ orientation gains for the optimization process. For our conditional scene synthesis module, we train our model on the same dataset excluding the input rooms used in our experiment. As the MatterPort3D dataset does not offer pose annotation, we use the rapid-annotation tool in \cite{keshavarzi2020scenegen} to label pose data within the scenes. 

Figure \ref{fig:3DExamples} shows the results of three sets of real-world captured rooms, each including rooms with different room sizes and functions. After extracting semantic labels of the objects (steps 2,3), the system performs mutual function optimization with functional semantics (step 3,4). Our proposed system is able to locate mutual walkable, sittable and workable functions in target rooms and align the physical environments to maximize the mutual functions. Furthermore, the system aims to complete the initialized synthetic rooms with the conditional scene synthesis process (steps 5,6).

\subsection{User Studies}
In a comparative user study, we aim to measure the participant's ability to find the maximum mutual functions between the rooms and compare it with the outcomes of our proposed mutual function system. We recruited 25 participants (m=10, f=15), which were skilled in 3D annotations to find mutual walkable and sittable functions of groups of rooms. We utilized 17 rooms of the MatterPort3D dataset which were organized in groups of three, and one group of two. We developed a 3D annotation application, which allowed participants to view all the rooms of the group in 3D, and annotate what they believed is the mutual areas between them. The tool also allowed the modification of annotated geometry after initial annotation. Participants were not aware that they were going to be compared to an automated system and were just asked to provide their best annotation skills for the task. Before data collection, the experiment operator demonstrated an example of how to use the annotator tool and answered questions on what is considered a mutual space. The data collection process from each participant took approximately 30 minutes, allocating 5 minutes to each room group for indicating mutual walkable and sittable spaces.

However, as anticipated, participants where not able to annotate 3 exactly similar areas in all three rooms. Therefore, in our analysis, we performed an extra step of aligning the user annotated spaces in a brute-force search process. The polygons are centred in a mutual point, and an exhaustive search is conducted between all possible orientations of the polygons to calculate the maximum intersection between them. We denote the maximum intersection as $M_U$ and further compare to $M_W$ and $M_S$ predicted by our system. The optimization implementation of the system were similar to what was described in Section \ref{sec:sceneGen_exp}.

Figure \ref{fig:userStudies} shows the top 3 highest mutual area classification task (out of 25) for walkable spaces performed by the users in green, compared to the system's calculation $M_S$ illustrated in blue for all six groups. As seen in the figure, the automated system clearly outperforms user performance in this classification task. A common technique that was observed is that most human annotators aimed to start with the smaller room and try to find corresponding spaces in the other rooms. This of course requires numerous editing attempts for the mutual space geometry to be modified accordingly.

\begin{figure}[t]
  \includegraphics[width=1\columnwidth]{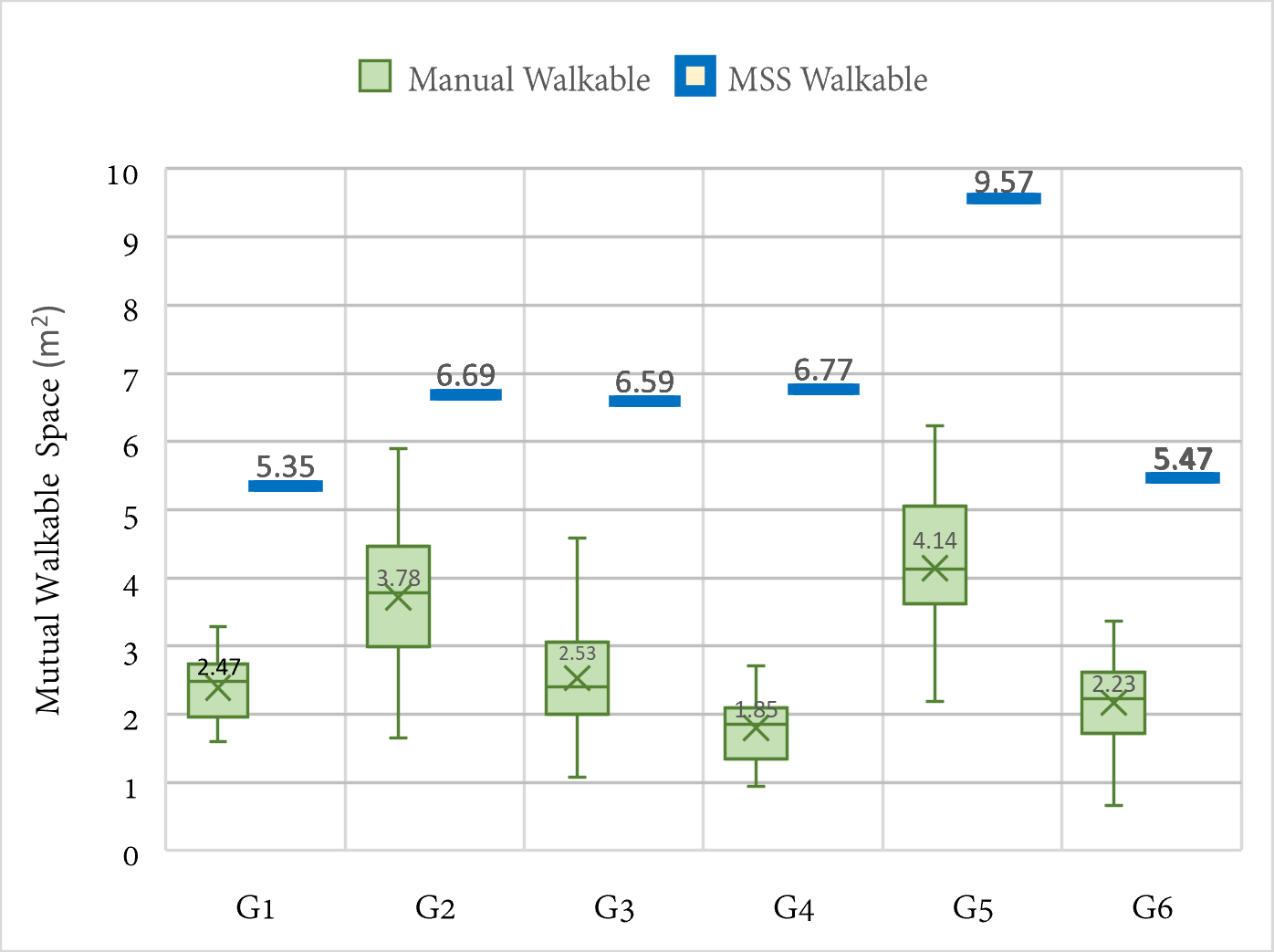}
    \includegraphics[width=1\columnwidth]{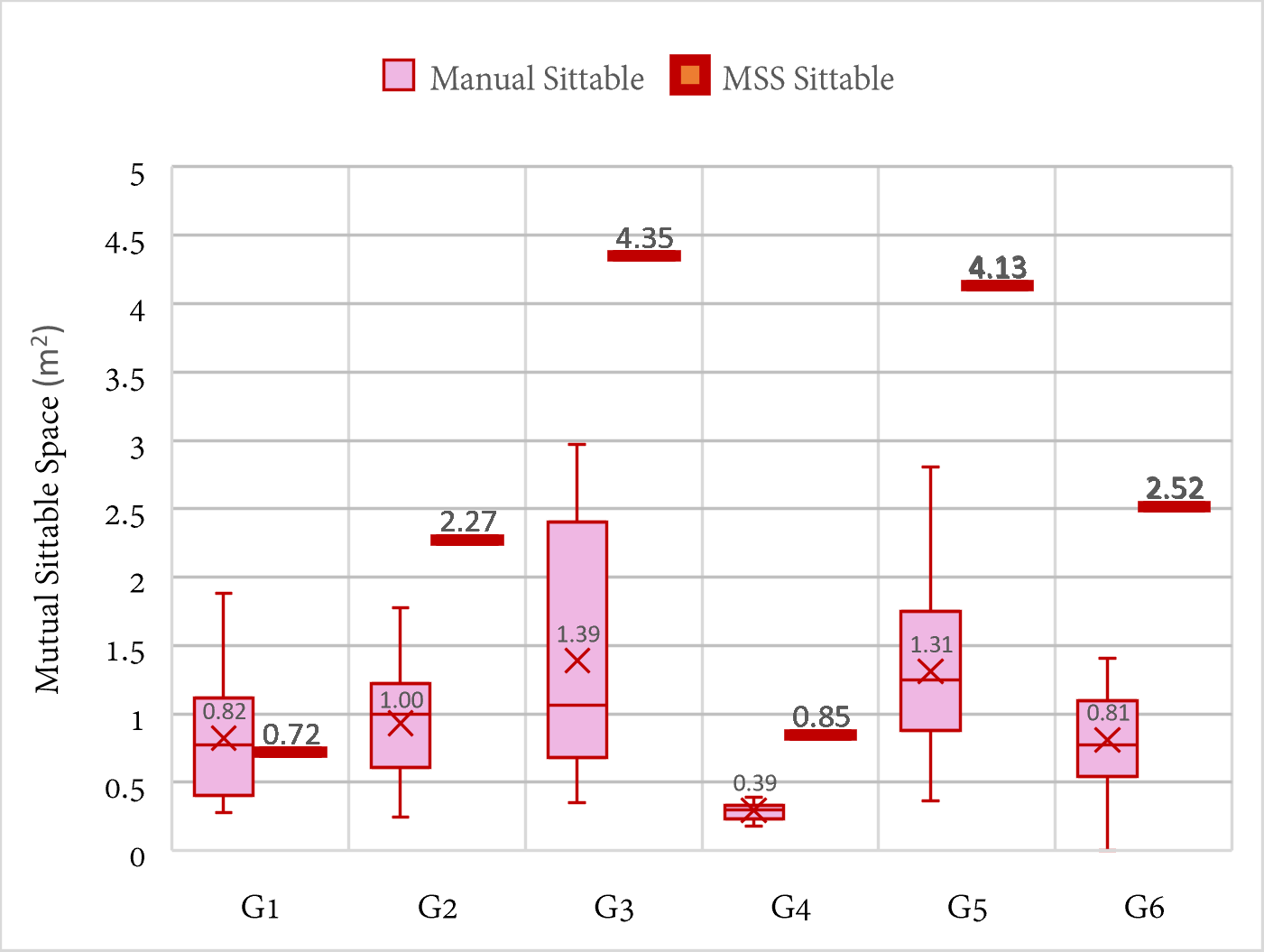}
  \caption{Comparison between manual annotations ($M_U$) and our MSS system for indicating mutual walkable ($M_W$) and sittable spaces ($M_S$) in 6 Matterport3D room groups}~\label{fig:userStudyChart}
\end{figure}

Figure \ref{fig:userStudyChart} shows a numerical comparison of the mutual area indication task between human-annotators ($M_U$) and our mutual scene synthesis system (MSS) for walkable and sittable spaces. For each room group, we plot a whisker-plot to visualize the distribution of $M_U$ for all participants, while a thick line represents MSS calculation. As seen in the figure, our system significantly finds larger areas of mutual spaces than human annotators with an average increase of 58.68\% for walkable spaces and 56.00\% average increase for sittable spaces. 

\section{Discussions}

As presented in our results from real-world captured room examples, furniture topology in the resulting synthetic scene often corresponds to objects present in physical environments. For instance in Figure \ref{fig:3DExamples}, in Group A, chairs and tables correspond to the location of office space $R_1$, while the storage space can also be attributed to $R_3$. All rooms hold part of their desk space as a mutual workable space. In Group B, a mutual sittable space is extracted from the area attributed to the bed in the bedrooms and a the larger sofa in the living room. In Group C, we see how small spaces such as the bedroom $R_7$ can also contribute to generating plausible spaces using our system. While the mutual spaces are considered limited, yet the resulting synthetic scene has utilized non-colliding functions from $R_8$ in its procedural generation before completing the space with additional contextual furniture.

The results of the user study show that regardless of annotation accuracy and performance, manually aligning mutual spaces is considered a challenging task for individuals. Identifying an acceptable boundary, and checking whether all rooms comply to the defined geometry can take multiple iterations of modifications. Such a process is time-consuming, and can be potentially difficult to execute for novice users in spatial computing platforms. The task may become more challenging in the event that privacy concerns are considered, preventing users to view other participant spaces during the telepresence setup. In absence of mutual space generation systems, users would need to communicate with each other to find suitable conditions that would address all spatial needs.

In the user study, there are a number of exceptional instances that participants classify larger walkable spaces than our automated system. This is because (a) we do not cross-validate the participants annotation as we consider any walkable or sittable area defined by the user to be correctly annotated (b) the system uses annotated labels from the dataset which are also annotated by humans and prone to error. However, since manually defining constant shapes between all rooms using our tool was seen as challenging, the actual mutual space post-processed by an exhaustive search module resulted in a significantly smaller area than each annotated room. Part of the classification inconsistencies could be attributed to the limitations in the annotation tool (eg. duplicating the annotation from one room to the other was not possible), hence, enhancing workflows to improve user classification could have changed the outcome of the experiment.

For the conditional scene synthesis module, a major challenge when relying on learning-based methods is that they are heavily biased towards the data. While the initial phase of our proposed scene augmentation module integrates a procedural approach, the final steps include populating the scene with additional furniture learned for scene priors. Real-world spaces are not generally designed for hosting virtual users. Hence, defining which scenes from the dataset are suitable for a meeting setting can be a challenging process. Models can be trained to filter room functions such as meeting-room spaces and offices-space, however, many spaces cannot be specifically classified to hold a single room functionality. For instance, a captured space from a studio or a dining room can serve as multiple functions. Another limitation of real-world datasets is their low label accuracy due to the labour intensive manual annotation process.  

\section{Limitations and Future Wok}
Our proposed system comes with a number limitations and failure cases. In scenarios with a large number of participants, the mutual space optimization module may fail to locate mutual function spaces that are present in all participants' spaces. In such cases, the current system relies on the procedural module to initialize a virtual scene using non-colliding functions. If this step is also implausible due to the furniture arrangement of target rooms, the system can fail in generating a mutual space. An alternative mechanism is to locate mutual spaces for subgroups, and initialize the scene augmentation process from output of the subgroup mutual space. This, however, would significantly increase the complexity of the optimization, as the system would need to initially search for the best subset of rooms. 

Another limitation of our workflow is seen in the iterative scene augmentation process for completing the virtual scene. The layout is dependent on the order of the object placement and does not calculate all possible permutations of the possible arrangements. Integrating robust floorplanning techniques to optimize the arrangement of all room elements present in the scene can be explored as future work. Moreover, our current framework is also limited to planar spaces, thus input rooms that include ramps or stairs cannot be processed correctly within the workflow. Expanding the formulation of our work to address building-scale environments with various elevations and floors can be possible next steps of this line of research.

 Conducting additional user experiments through developing mixed reality prototypes can help identify the challenges of such system from a user standpoint. Exploring effective techniques for users to interact with a synthetic scene generator, while allowing them to modify and adjust the output of such systems can be studied. Moreover, usability studies can be performed to identify best strategies for user-in-the-loop input during the multi-function optimization process. Finally, improving the framework to address scenarios with more than one person present in each space can be explored. 

\section{Conclusion}
In this paper we have presented a method for synthesizing a virtual environment for telepresence settings which corresponds to the spatial arrangement of the participants' physical local environments. Our method aims to calculate the maximum mutual walkable, sittable and workable spaces between users, allowing the synthesized virtual scene to hold areas of mutual ground for efficient virtual interaction. We utilize state-of-the-art scene synthesis methods to populate the virtual room with objects that hold topological and functional relationships with elements of the scene. We extend the scene augmentation process by introducing a conditional mechanism, allowing virtual objects to position themselves close to objects with same functionalities in the physical environment.

Our experiments demonstrate how our proposed mutual scene synthesis method works in action by using real-world captured rooms in the MatterPort3D dataset as input to the system. We show that using our method, meaningful spaces suitable for meeting spaces are synthetically generated while holding functional mutual areas for users to utilize. Furthermore, by performing a series of user studies to compare task performance between manual and automated mutual space classification, we show our proposed time is able to locate significantly larger mutual spaces in a fraction of the time. We expect manually identifying mutual spaces to come with additional difficulties when conducted in mixed reality, while privacy concerns of sharing participant room layouts are to be considered. Hence, an automated system to generate synthetic spaces can potentially facilitate the adaption of mixed reality telepresence platforms.

\acknowledgments{
This research was conducted during Mohammad Keshavarzi's research internship at Reality Labs Research (RLR), Meta. We thank the larger team at RLR for providing technical feedback and for their assistance during the comparative user studies.
}

\bibliographystyle{abbrv-doi}

\bibliography{template}
\end{document}